\documentclass[%
 reprint,superscriptaddress,amsmath,amssymb,
 aps,nofootinbib]{revtex4-1}
\usepackage{dsfont}
\usepackage{graphicx}

\usepackage{dcolumn}
\usepackage{bm}
\usepackage{color}
\usepackage{xcolor}
\usepackage{epsfig}
\usepackage{soul}
\usepackage[colorlinks=true,urlcolor=brown,linkcolor=blue,citecolor=magenta]{hyperref}
\usepackage{natbib}
\usepackage{float}
\usepackage{footmisc}
\usepackage{siunitx}
\usepackage{soul}

\definecolor{lightred}{rgb}{1,0.5,0.5}
\definecolor{lightgreen}{rgb}{0.5,1,0.5}
\definecolor{lightblue}{rgb}{0.5,0.5,1}
\definecolor{lightcyan}{rgb}{0.5,0.75,0.75}
\definecolor{lightmagenta}{rgb}{0.75,0.5,0.75}
\definecolor{customgreen}{rgb}{0.494,1,0.502}

\newcommand{\GeV}{\mathinner{\mathrm{GeV}}}
\newcommand{\TeV}{\mathinner{\mathrm{TeV}}}

\begin{document}

\title{Magnetogenesis with gravitational waves and primordial black hole dark matter}

\author{Shyam Balaji} 
\email{shyam.balaji@kcl.ac.uk}
\affiliation{Physics Department, King’s College London, Strand, London, WC2R 2LS, United Kingdom}

\author{Malcolm Fairbairn} 
\email{ malcolm.fairbairn@kcl.ac.uk}
\affiliation{Physics Department, King’s College London, Strand, London, WC2R 2LS, United Kingdom}

\author{María Olalla Olea-Romacho\vspace{3mm} }
\email{mariaolalla.olearomacho@phys.ens.fr}
\affiliation{Laboratoire de physique de l'École normale supérieure, ENS, Université PSL, CNRS, Sorbonne Université, Université Paris Cité, F-75005 Paris, France}

\begin{abstract}
Strongly supercooled first order phase transitions (FOPTs) can produce primordial black hole (PBH) dark matter (DM) along with observable gravitational waves (GWs) from bubble collisions. Such FOPTs may also produce coherent magnetic fields generated by bubble collisions and by turbulence in the primordial plasma. Here we find that the requirement for PBH DM can produce large primordial magnetic fields which subsequently yield intergalactic magnetic fields in the present universe (with magnitude $\lesssim 20$ pG across coherence length scales of $\simeq 0.001$-$0.01$ Mpc, assuming maximally helical magnetic fields) that easily exceed lower bounds from blazar observations. We follow a largely model independent approach and highlight the possibility of producing DM and observable multi-messenger magnetic fields and GW signals visible in next generation experiments.
\end{abstract}

\maketitle

\section{Introduction}

Dark matter (DM), remains one of the most important open questions in physics and cosmology. One popular possibility is for all of the DM to be comprised of primordial black holes
(PBHs)~\cite{Zeldovich:1967lct,Hawking:1971ei, Carr:1974nx,Carr:2020gox} if their masses are in the asteroid-mass range \cite{Balaji:2022rsy,Qin:2023lgo}. It has been demonstrated that PBHs can be copiously generated in the supercooling regime of a first order phase transition (FOPT) when the energy density of the universe is
controlled by the latent heat of a phase transition~\cite{Kodama:1982sf, Hsu:1990fg}. The latent
heat behaves as a cosmological constant which drives the expansion of the universe until the transition ends and the energy is
transformed into radiation when bubbles form and merge, subsequently reheating the plasma. Stochastically late-nucleated Hubble patches lead to horizon-scale overdensities that collapse into PBHs upon completion of the transition.

  Primordial magnetogenesis has also been studied in relation to FOPTs, such as e.g.~the electroweak (EW) phase transition~\cite{Vachaspati:1991nm, Ellis:2019tjf, Olea-Romacho:2023rhh} or the QCD phase transition~\cite{Sigl:1996dm, Tevzadze:2012kk}. Notably, the idea of magnetogenesis during a first-order electroweak phase transition was first proposed in Ref.~\cite{Vachaspati:1991nm}. In this scenario, magnetic fields are created by EW sphaleron decays~\cite{Vachaspati:2001nb, Copi:2008he, Chu:2011tx}. As the bubbles formed during the EW phase transition grow, collide and merge, they cause the primordial plasma to move at high Reynolds
number, and the magnetic fields enter a state of magnetohydrodynamic (MHD) turbulence~\cite{Witten:1984rs, Hogan:1986qda, Kamionkowski:1993fg, Brandenburg:1996fc, Christensson:2000sp, Kahniashvili:2010gp, Brandenburg:2017neh}. 

Coherent intergalactic magnetic fields (IGMFs) are indirectly suggested by blazar observations~\cite{MAGIC:2022piy, HESS:2023zwb, Neronov:2010gir}. The origin of these IGMFs is a long-standing mystery, with two main possibilities: astrophysical or cosmological. An example of astrophysical sources is weak initial magnetic fields from local effects in astrophysical objects (e.g. the Biermann battery mechanism~\cite{PhysRev.82.863}) that are later enhanced by dynamo effects~\cite{AlvesBatista:2021sln}, which can create such long-correlated magnetic fields. Cosmological explanations from FOPTs can also easily accommodate magnetic fields with very large coherence lengths. Hence, potential sources from early universe cosmological processes such as FOPTs are very attractive. 

In this paper, we will use model-independent results for the inverse duration of the FOPT duration and reheating temperature that produce PBHs as all DM, we then compute the corresponding IGMFs in the present universe and compare them with blazar lower bounds. Strong FOPTs also produce a stochastic background of GWs~\cite{Witten:1984rs,Hogan:1986qda,Roshan:2024qnv}. By enforcing the production of PBH DM in the transition, we also compute the corresponding gravitational wave (GW) signal from the bubble collisions, leading to observable signals in upcoming GW interferometers well above stochastic astrophysical foregrounds~\cite{Baldes:2023rqv}. In this way, we highlight the relation between mulitmessenger cosmological signals of shared origin in two disparate channels, magnetic fields and GWs, whilst simultaneously providing an explanation for DM. We also briefly remark on a possible model realisation via the classically conformal $U(1)_{B-L}$ model for completeness.

This paper is structured as follows: in Sec.~\ref{sec:pbhsbfieldsGWs}, we provide (i) a brief overview of PBH formation from FOPTs, (ii) we show how to compute the resulting primordial magnetic field and coherence length and outline IGMF bounds from blazars and (iii) we outline how to calculate GW spectrum and verify observability in various experiments. In Sec.~\ref{sec:results}, we discuss the main results of this work and finally we conclude in Sec.~\ref{sec:conclusion}. 

\section{Primordial black holes, magnetic fields and gravitational waves}
\label{sec:pbhsbfieldsGWs}
\subsection{Primordial black hole formation}
\label{sec:PBH_formation}
In a FOPT, the universe can change
from a metastable symmetric vacuum to a symmetry-breaking vacuum,
through a process of bubble nucleation, growth and merging.
Supercooling occurs when the universe remains in the metastable phase for an extended period of time, such that the expansion of the universe is largely driven by the false vacuum energy rather than radiation.
The vacuum energy inflates the universe until it converts to radiation upon bubble coalescence, ultimately reheating the plasma to a temperature $T_{\rm reh}$.
Given the stochastic nature of the phase transition, PBHs can form in supercooled cosmological phase transitions. For the average background Hubble patch, nucleation occurs at the cosmic time $\tau_{\rm nuc}$. Causal regions (labelled by $i$) that nucleate at a time $t_{\rm nuc}^i$ later than $\tau_{\rm nuc}$ remain vacuum dominated for a longer period. The false vacuum energy does not change with the expansion of the universe, while the energy density of both the bubble walls and the radiation decreases rapidly.  For this reason, late-nucleating patches quickly become overdense with respect to the surrounding background patches, and if the overdensity  exceeds a certain threshold $\delta_c$, these regions may collapse into PBHs. 
We use the density contrast in radiation to determine the collapse condition as
\begin{equation}
	\delta(t) \equiv \frac{ \rho_{r}^{\rm late}(t; t_\textrm{{nuc}}^i) - \rho_{r}^{\rm bkg}(t) }{ \rho_{r}^{\rm bkg}(t) } > \delta_c,
\end{equation}
where $\rho_{r}^{\rm late}(t; t_\textrm{{nuc}}^i)$ is the radiation density in a region where nucleation is delayed to $t_\textrm{{nuc}}^i$ and $\rho_{r}^{\rm bkg}(t)$ is the radiation energy density of a background Hubble patch at cosmic time $t$.
The density contrast peaks shortly after the late patch percolation, as the energy density in the background patch begins diluting slightly earlier, while the late patch keeps a constant energy density due to the vacuum energy density.

The inverse timescale of the transition is given by
	\begin{equation}
	\frac{\beta}{H} \equiv \frac{1}{H\Gamma_{\rm bub}}\frac{d\Gamma_{\rm bub}}{dt},
	\end{equation}
where $\Gamma_{\rm bub}$ is the bubble nucleation rate per unit volume and $H$ is the Hubble rate during the phase transition.\footnote{In the following, we will use a first-order Taylor expansion to approximate the bubble nucleation rate near $\tau_{\rm nuc}$. In Ref.~\cite{Baldes:2023rqv}, it was shown that the approximate form yields accurate predictions around the nucleation time.}
The Hubble rate in the false vacuum reads,
	\begin{equation}
	 H \equiv H_{\rm false} = \sqrt{ \frac{8 \pi}{3 M_{\rm Pl}^2 } \rho_{\rm vac} } ,
	\end{equation}
where $\rho_{\rm vac}$ is the vacuum energy density just before the onset of the phase transition. 

In order to quantify the temperature at the moment that the phase transition completes, and assuming that the universe rapidly reheats to the reheating temperature at a rate much greater than the expansion rate of the universe. We define the reheating temperature as 
\begin{equation}
	T_{\rm reh} =  \left( \frac{30 \rho_{\rm vac} }{g_{r}(T_\textrm{reh})\pi^{2}}  \right)^{1/4},
	\end{equation}
where $g_{r}$ are the radiation degrees-of-freedom. This temperature should be understood as the temperature in the false vacuum patch, which does not contain nucleated bubbles or partially reheated plasma.

For a monochromatic distribution, the mass of these PBHs is given by the energy inside the sound horizon at the time of the collapse~\cite{Gouttenoire:2023naa}
\begin{equation}
M_{\rm PBH}\left(T_\textrm{reh}\right) = M_\odot\left(\frac{20}{g_{r}(T_\textrm{reh})}\right)^{1/2}\left(\frac{140\,\textrm{MeV}}{T_\textrm{reh}}\right)^2,
\label{eq:pbhmasses}
\end{equation}
whereas the PBH abundance normalised to the DM relic abundance reads~\cite{Gouttenoire:2023naa}
\begin{align}
f_\textrm{PBH}\left(\beta/H,T_\textrm{reh}\right)=\frac{e^{-a_1 (\beta/H)_{\ast}^{a_2} \left(1+\delta_c\right)^{a_3 (\beta/H)_{\ast}}}}{2.2\times 10^{-8}} \frac{T_\textrm{reh}}{0.14}.
\label{eq:abundance}
\end{align}
Here,
$a_1=0.56468$, $a_2=1.266$ and $a_3 = 0.6639$, and $\beta/H$ is evaluated at the percolation time, yielding
\begin{align}
    (\beta/H)_*=\frac{a_2 W_0\left[\frac{a_3}{a_2}\left(\frac{19.5983+\log T_\textrm{reh}}{a_1}\right)^{\frac{1}{a_2}}\log(1+\delta_c)\right]}{a_3 \log(1+\delta_c)},
    \label{eq:beta_H_star}
\end{align}

where $W_0$ is the zeroth order Lambert $W$ function. Calculations, based on full general relativistic simulations, often suggest a range of 0.4 to 0.66 for $\delta_c$ in the context of overdensities from inflation re-entering the Hubble horizon~\cite{Carr:1975qj, Shibata:1999zs, Musco:2004ak, Harada:2013epa, Musco:2018rwt,Musco:2020jjb, Germani:2018jgr,Escriva:2019phb, Stamou:2023vxu}. 
Deviations from these assumptions are expected in the phase transition scenario. However, in this study, we align with previous approaches~\cite{Gouttenoire:2023naa, Baldes:2023rqv}, and choose $\delta_c=0.45$ to enable  comparison. The precise value of $\delta_c$ impacts $f_{\rm PBH}=1$, yet the strong magnetic field generated remain insensitive to this choice.

\subsubsection{Experimental constraints for PBHs as the dark matter}
Observationally, $f_{\rm PBH}=1$ is allowed for nearly monochromatic distributions of PBHs with masses falling within what is known as the asteroid-mass window~\cite{Carr:2021bzv}
\begin{equation}
10^{-16}M_{\odot} \lesssim M_{\rm PBH} \lesssim 10^{-10}M_{\odot}.
\label{eq:windowallowed}
\end{equation}
Here the lower bound is set by the particle flux from Hawking evaporation, which would affect the CMB~\cite{Carr:2009jm} or be detected  by the Voyager~\cite{Boudaud:2018hqb} probe. The upper bound is set by microlensing measurements~\cite{Niikura:2017zjd}. According to Eq.~(\ref{eq:pbhmasses}), 
the range of reheating temperatures that gives rise to 
PBHs with masses within the asteroid-mass window is 
\begin{equation}
10 \TeV \lesssim T_{\rm reh} \lesssim 10^4 \TeV.
\end{equation}
To get $f_{\rm PBH}=1$ in this temperature range, we use Eq.~(\ref{eq:beta_H_star}).

\subsection{Primordial magnetic fields}
\label{sec:PMF}

\subsubsection{Magnetic field spectrum today}

The initial conditions of the magnetic field and the plasma, which are not well understood, have a significant impact on how MHD turbulence decays. For instance, the initial seed field magnetic helicity is one of the factors that influences the decay of MHD turbulence~\cite{Banerjee:2004df}. The magnetic helicity is a quantity that describes how twisted and linked the magnetic field lines are. It is given by $\langle \mathbf{A} \cdot \mathbf{B} \rangle$, where $\mathbf{B}=\nabla \times \mathbf{A}$, where $\mathbf{A}$ and $\mathbf{B}$ are the usual vector potential and magnetic field respectively.
The magnetic helicity is nearly conserved in a highly conductive medium. This implies that a field with maximum helicity has to increase its correlation length as it loses magnetic energy, which leads to an inverse cascade of magnetic energy where the energy moves from smaller scales to larger scales, creating coherent magnetic structures at scales much bigger than the ones where the energy was initially injected. Therefore, this process could have been very important for the survival and evolution of primordial magnetic fields, which would be  correlated at very large length scales today.

The MHD turbulence for maximally helical magnetic fields decays as a power law in conformal time, $\eta$, with
the following relations for the magnetic energy
and correlation length during the radiation-dominated epoch~\cite{PhysRevLett.83.2195}
\begin{equation}
B \sim \eta^{-1/3} \, \, \mathrm{and} \, \, \lambda \sim \eta^{2/3}.
\end{equation}

It has been shown by numerical simulations that magnetic fields with zero or negligible net helicity can also experience an inverse transfer of magnetic energy when there is a plasma with initial kinetic helicity~\cite{Brandenburg:2017rnt}.\footnote{However, this has been challenged recently in Ref.~\cite{Armua:2022rvx}, where a weaker inverse cascade of magnetic energy for non-helical fields was found, compared
to previous studies in the literature.}  In this scenario, the magnetic energy and
correlation length, $\lambda$, scale as
\begin{equation}
    B \sim \eta^{-1/2} \,\, \, \mathrm{and} \,\, \, \lambda \sim \eta^{1/2} \, ,
\end{equation}
respectively, during the radiation-dominated epoch.
These scaling laws are valid during the epoch of radiation-domination, when the scale factor increases linearly with conformal time, i.e.~$a \sim \eta$. After recombination, the magnetic field decreases like radiation, i.e.~$B \sim a^{-2}$.
To express these two scenarios in a concise way, we introduce
the parameters
\begin{equation}
q_b = \frac{2}{b+3} \, \, \, \, \mathrm{and} \, \, \, \,  p_b = \frac{2}{b+3}(b+1) \, ,
\end{equation}
for the power-laws
\begin{equation}
    B \sim \eta^{- p_b / 2} \, \,\, \, \mathrm{and} \,\, \, \, \lambda \sim \eta^{q_b} \, ,
\end{equation}
where the cases $b=0$ and $b=1$ correspond to the maximally helical and non-helical scenarios described above, respectively.

The magnetic field energy density 
at percolation can be estimated by~\cite{Ellis:2019tjf, RoperPol:2023bqa}
\begin{equation}
\rho_{B,*}  = 0.1\frac{ \kappa_{\rm col} \alpha}{1+ \alpha} \rho_{*} \approx \frac{ \pi^2}{3} T_{\rm reh}^4 \approx 0.1 \, \rho_{\rm vac}
\label{eq:mag_energy_dens}
\end{equation}
Here, $\rho_{*}=\frac{3M_{\rm Pl}^2}{8\pi}H_{*}^2 = \frac{g_{r}(T_\textrm{reh})\pi^2}{30} T_{\rm reh}^4 \approx \rho_{\rm vac}$ is the total energy density at the percolation temperature. Here $\kappa_{\rm col}$ is the fraction of the released vacuum energy that is used in accelerating the bubbles. The
efficiency for converting bulk fluid motion of the plasma into magnetic fields via
MHD turbulence was assumed to be of $10 \%$~\cite{Kahniashvili:2009qi,Durrer:2013pga, Brandenburg:2017neh}. For supercooled phase transitions, we can safely assume $\alpha \rightarrow \infty$ and $\kappa_{\rm col} = 1$, leading to the approximation in Eq.~\eqref{eq:mag_energy_dens}.

The magnetic field spectrum today can be computed as~\cite{Ellis:2019tjf}
\begin{align}
\label{eq:spectrum}
&B_0(\lambda) \equiv B(\lambda,t_0) = \\ &\left(\frac{a_{\rm reh}}{a_{\rm rec}}\right)^{p_b/2} \left(\frac{a_{\rm reh}}{a_0}\right)^2 \sqrt{\frac{17}{10}\,\rho_{B,*}} \,
\begin{cases}
\left(\frac{\lambda}{\lambda_0}\right)^{-5/2}  & \mbox{ for }~~ \lambda\geq\lambda_0 \, \\    \left(\frac{\lambda}{\lambda_0}\right)^{1/3} & \mbox{ for }~~ \lambda < \lambda_0 \,, \nonumber
\end{cases}
\end{align}
which assumes a power-law spectrum for the magnetic field strength, with a spectral index of $n=2$ at large scales. 
Here $\lambda_0$ denotes the field coherence scale redshifted to today~\cite{Ellis:2019tjf}
\begin{equation}
\label{eq:l0}
\lambda_0 \equiv \lambda_B(t_0) = \left(\frac{a_{\rm rec}}{a_{\rm reh}}\right)^{q_b} \left(\frac{a_0}{a_{\rm reh}}\right) \lambda_*, \,
\end{equation}
where the initial correlation length $\lambda_{*}$ is given by the bubble size at percolation $R_{*}$~\cite{Caprini:2019egz}
\begin{equation}
\lambda_{*} = R_{*} = \frac{(8 \pi)^{1/3}}{H_{*}} \left( \frac{\beta}{H} \right)^{-1},
\end{equation}
where, for the case of supercooled phase transitions, a bubble wall velocity of $v_{w}=1$ was assumed. The peak value of the magnetic field spectrum is denoted as $B_{0} = B_{0}(\lambda_{0})$.
The redshift factors are computed as
\begin{align}
&\frac{a_{\rm reh}}{a_0} = 8 \times 10^{-14} \left( \frac{100}{g_{r}(T_{\rm reh})} \right)^{1/3} \left( \frac{1 \GeV}{T_{\rm reh}} \right) \, , \\
&\frac{a_{\rm reh}}{a_{\rm rec}} = 8 \times 10^{-11} \left( \frac{100}{g_{r}(T_{\rm reh})} \right)^{1/3} \left( \frac{1 \GeV}{T_{\rm reh}} \right) \, ,
\end{align}
where $g_{r}(T_{\rm reh})$ corresponds to the total number of relativistic degrees of freedom in entropy at the reheating  temperature $T_{\rm reh}$. For this work, we will assume $g_{r}(T_{\rm reh})=106.75$ as in the case of the Standard Model (SM). However, this value is model dependent and would have to be adjusted to reflect the dynamics of the UV complete theory.

If the transition is driven by dark scalars, coupled through a portal to the SM Higgs field, a small fraction of the bubble wall energy can be stored in the Higgs field bubble. We take into account this suppression by defining an efficiency parameter $\kappa_h$ which indicates how much energy has been transferred to the SM Higgs field.



\subsubsection{Experimental constraints from blazar emissions}

Blazars are active galactic nuclei jets that are roughly
pointed towards us, and they emit TeV $\gamma$-rays. These $\gamma$-rays can collide with photons from the extragalactic background, creating $e^{+}e^{-}$ pairs \cite{Balaji:2022wqn}. Then, these pairs can interact with photons from the CMB, producing $\gamma$-rays with energies in the GeV range. This in turn modifies the original spectrum of the blazars by decreasing the number of TeV $\gamma$-rays and increasing the number of GeV $\gamma$-rays~\cite{Vachaspati:2016xji,Vachaspati:2020blt}. In the presence of IGMF, the $e^{+}e^{-}$ pairs can be deflected by the Lorentz
force and, if the field is strong enough, the final GeV photons are no longer directed towards the observer. The absence of these photons has been used to set lower bounds on the strength of the IGMF.

We considered two recent analyses of blazar observations that measure the minimum strength of the IGMF. The first one, done by the MAGIC $\gamma$-ray observatory, sets a lower bound of $ B > 1.8 \times 10^{-17} \, \mathrm{G}$ for correlation lengths $\lambda \geq 0.2 \, \mathrm{Mpc}$~\cite{MAGIC:2022piy}. The second one, based on the data from the \textit{Fermi}-LAT and H.E.S.S.~collaborations, establishes a limit of $ B > 7.1 \times 10^{-16} \, \mathrm{G}$ for coherence lengths of $\lambda \geq 1 \, \mathrm{Mpc}$, assuming a blazar duty cycle of $t_d = 10 \, \mathrm{yrs}$~\cite{HESS:2023zwb}. 

\subsection{Gravitational waves}
\label{sec:GWs}
We use the results from hybrid simulations~\cite{Lewicki:2020azd} to estimate the anisotropic stress from a bubble collision, where the authors first simulated the collision in one spatial dimension, then used the result as a source at points at which walls collide in a three-dimensional lattice simulation  with thin walls. This way, the effects of scalar gradients and gauge field production during the collision can be captured by the lower dimensional simulation. The spectrum is given by
	\begin{equation}
	h^{2}\Omega_{\rm gw}(f) = 5.10 \times 10^{-9} \left( \frac{ 100 }{ g_{r}(T_\textrm{reh}) } \right)^{1/3} \left( \frac{ 10 }{ (\beta/H)_{*} } \right)^{2}  S_{\rm hy}(f),
	\end{equation} 
where the shape is
	\begin{equation}
	S_{\rm hy}(f) = \frac{ 695 }{ \left[ 2.41 \left( \frac{f}{\tilde{f}_{\rm hy}} \right)^{-0.557} + \left(  \frac{f}{\tilde{f}_{\rm hy}} \right)^{0.574} \right]^{4.20} }
	\end{equation}
and peak frequency
\begin{equation}
	\tilde{f}_{\rm hy} =  22 \; \mathrm{mHz} \, \left( \frac{ g_{r}(T_\textrm{reh}) }{ 100 } \right)^{1/6} \left(  \frac{ (\beta/H)_{*} }{ 10 } \right)   \, \left( \frac{ T_{\rm reh} }{ 10^{2} \; \mathrm{TeV} } \right).
	\end{equation}
The signal-to-noise ratio is given by~\cite{Allen:1997ad,Kudoh:2005as,Thrane:2013oya,Caprini:2019pxz,Brzeminski:2022haa}
	\begin{equation}
	\mathrm{SNR} = \sqrt{t_{\rm obs} \int \left(\frac{\Omega_{\rm gw}^2}{\Omega_{\rm sens}^2 + 2\Omega_{\rm gw}\Omega_{\rm sens}+	2\Omega_{\rm gw}^2}\right) df},
	\label{eq:snr}
	\end{equation}
where $t_{\rm obs}$ is the observation time and $\Omega_{\rm sens}$ the sensitivity of the interferometer. The predictions for the GWs produced by the supercooled FOPT that we consider yield compelling phenomenology in the observational window of Laser Interferometer Space Antenna (LISA), Einstein Telescope (ET), Cosmic Explorer (CE) and DECIGO (Deci-hertz Interferometer Gravitational wave Observatory). At higher frequency, we have a band that would be observable with the ET~\cite{Maggiore:2019uih}, which would be sensitive to the range $10$--$10^3$ Hz. In between Square Kilometre Array \cite{Balaji:2023ehk} and ET frequencies we expect LISA \cite{Barausse:2020rsu} and DECIGO \cite{Yagi:2011wg,Kawamura:2020pcg,Balaji:2022dbi} to be applicable.
In order to compare theoretical predictions with experimental projections, we use the recently derived future sensitivity curves for these GW limits using the experimental design specification along with the peak integrated sensitivity curves computed in Ref.~\cite{Schmitz:2020syl} and assume an observation time of 1 yr.

\section{Results}
\label{sec:results}
Assuming a supercooled phase transition that saturates the DM content of the universe as PBHs, we can compute the properties of the primordial magnetic fields created during the transition, such as their strength and coherence length. We can also evaluate the GW signal generated in this transition.
\begin{figure}[h]
\centering
\includegraphics[width=1\linewidth]{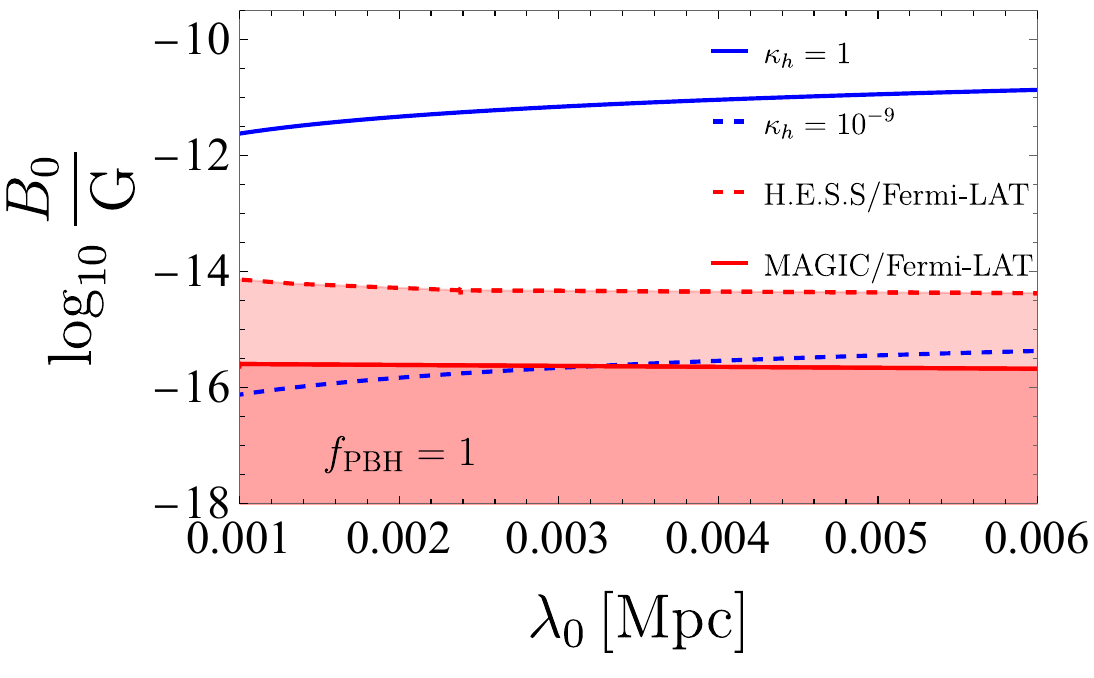}
\caption{Magnetic field peak value (blue line) in the present universe $B_0$ vs the coherence length $\lambda_0$ for a reheating temperature of $T_\textrm{reh} \in [10^4,10^7]$ GeV, $b=0$ and $f_\textrm{pbh}=1$. The solid red line corresponds to the lower bound on the magnetic field set by Ref.~\cite{MAGIC:2022piy}, whereas the dashed red line corresponds to that set by Ref.~\cite{HESS:2023zwb} for a duty cycle of $t_d = 10 \, \mathrm{yrs}$.}
\label{fig:B0l0}
\end{figure}

In Fig.~\ref{fig:B0l0}, we show the peak magnetic field amplitude, $B_0$, computed using \eqref{eq:spectrum}, for a maximally-helical magnetic field ($b=0$), as a function of the corresponding coherence length $\lambda_0$ at the peak magnetic field value, from~\eqref{eq:l0}, spanning over reheating temperatures of $[10^4,10^7] \GeV$.  The coherence length scales for combinations of reheating temperatures and $\beta/H$ values that produce $f_\textrm{pbh}=1$ lie in a relatively narrow window of $6\times 10^{-4}$ to $ 7\times 10^{-3}\, \rm Mpc$. Across this range, the magnetic field amplitude varies from around 1-15 pG. In the same figure, we show the peak magnetic field for an efficiency factor of $\kappa_h=10^{-9}$ (note $B_0$ scales like $\sqrt{\kappa_h}$) to illustrate the possible effect of bubble wall energy being stored in the SM Higgs field if the transition were driven by dark scalar through a portal to the Higgs field. 
For strongly supercooled transitions, $\kappa_h=10^{-9}$ is an estimate of the possible suppression inspired in scale-invariant extensions of the SM~\cite{Ellis:2019tjf} (see Sec.~\ref{sec:b-lmodel}).
The red solid line indicates the lower bound on $B_0$ from
MAGIC and \textit{Fermi}-LAT observations at the corresponding coherence length $\lambda_0$, whereas the dashed line corresponds to the one based on data from the \textit{Fermi}-LAT and H.E.S.S.~collaborations. In the absence of suppression ($\kappa_h=1$), the values of $B_0$ could explain the blazar observations, since the predicted values of $B_0$ easily exceed the lower bounds set by the experiments. Taking into account illustrative examples for the suppression featuring values of $\kappa_h$ as low as $10^{-9}$, we obtain values of $B_0$ that can exceed the lower bound set by MAGIC and \textit{Fermi}-LAT but are too weak to surpass the limits established by H.E.S.S. and \textit{Fermi}-LAT over most of the range of coherence lengths. However, the values for $B_0$ are still above the most conservative blazar bound set by MAGIC and \textit{Fermi}-LAT over coherence lengths of $\simeq 3.4 \times 10^{-3}$ Mpc.

We note that in the case of a non-helical magnetic field ($b=1$), the peak magnetic field gets suppressed by more than two orders of magnitude and lies in the range $2\times 10^{-3}$-$7\times 10^{-2}$ pG over smaller coherence length scales of $\simeq 9\times10^{-7}$-$3\times10^{-5}$ Mpc. This is of significantly less experimental interest, so we focus our discussion on the more relevant, maximally helical case. The maximally helical case is also phenomenologically interesting since a cosmological origin is plausible for correlated magnetic fields over large scales. Such mechanisms frequently involve an inverse cascade process that requires an initial net magnetic helicity. This is because the correlation length of the fields generated in the early universe cannot exceed the causal region at that time, so a way to transfer the magnetic energy to larger scales is needed. The inverse cascade, which occurs for helical fields, is a natural candidate for this. In the following we choose to disregard the case of non-helical magnetic fields ($b=1$), as the magnetic field amplitudes in this scenario are relatively suppressed. 

\begin{figure}[h]
\includegraphics[width=1\linewidth]{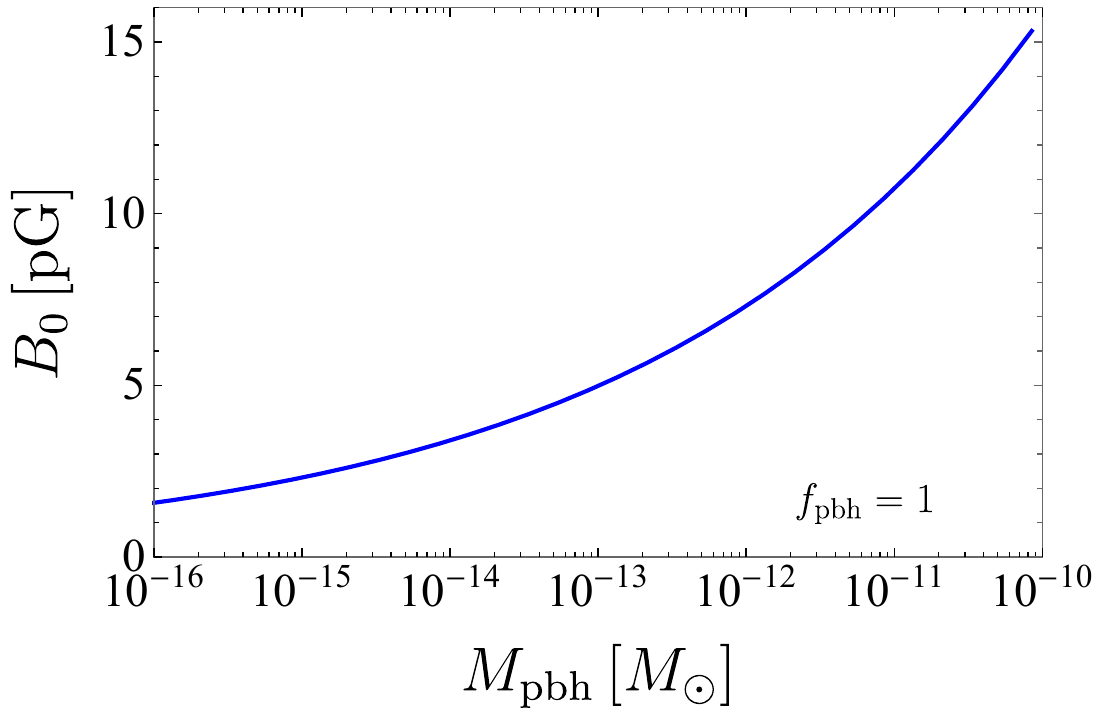}
\includegraphics[width=1\linewidth]{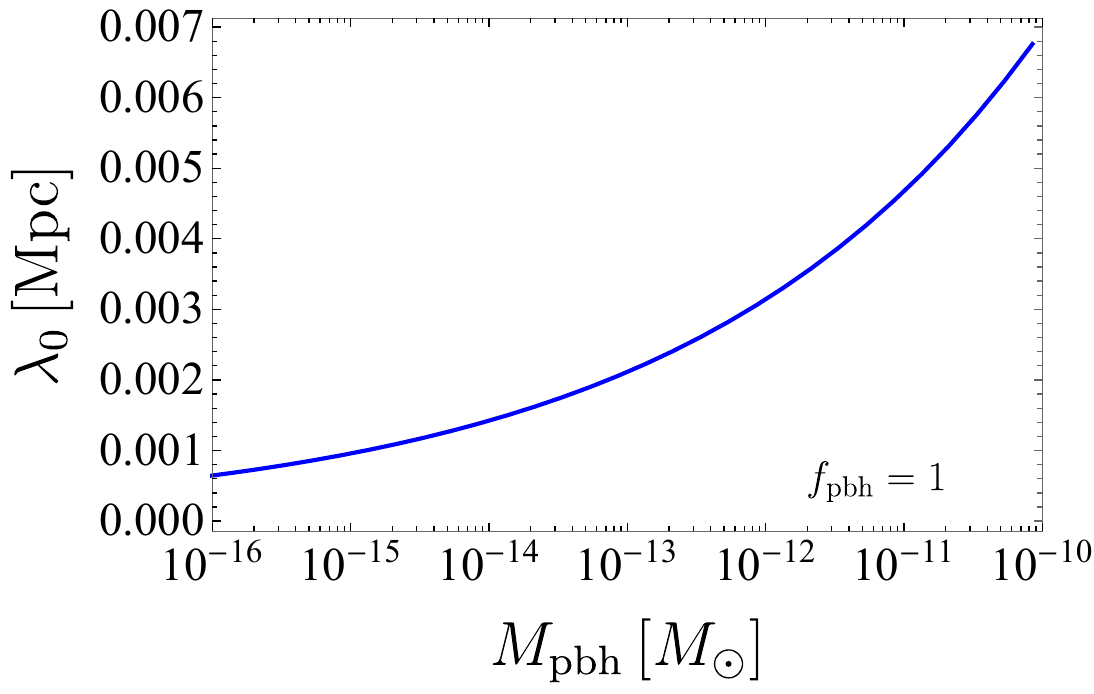}
\caption{Magnetic field peak value in the present universe $B_0$ vs the primordial black hole mass (top panel). Coherence length $\lambda_0$ vs the primordial black hole mass (bottom panel) is also shown. In both cases, we consider a reheating temperature of $T_\textrm{reh} \in [10^4,10^7]$ GeV, $b=0$ and $f_\textrm{pbh}=1$.}
\label{fig:B0l0mbph}
\end{figure}

We also plot the coherence length and magnetic field amplitude as a function of the PBH DM mass, as shown in Fig.~\ref{fig:B0l0mbph}. In the top and bottom panels we show the peak magnetic field amplitude and coherence length $\lambda_0$ as a function of the PBH mass respectively, computed using \eqref{eq:pbhmasses}, which spans the aforementioned asteroid mass range of $[10^{-16},10^{-10}]M_\odot$ for the reheating temperatures considered. PBHs as DM are correlated with coherent magnetic fields at length scales from around $0.001$ to $0.01 \, \rm Mpc$.

\begin{figure}[h]
\includegraphics[width=1\linewidth]{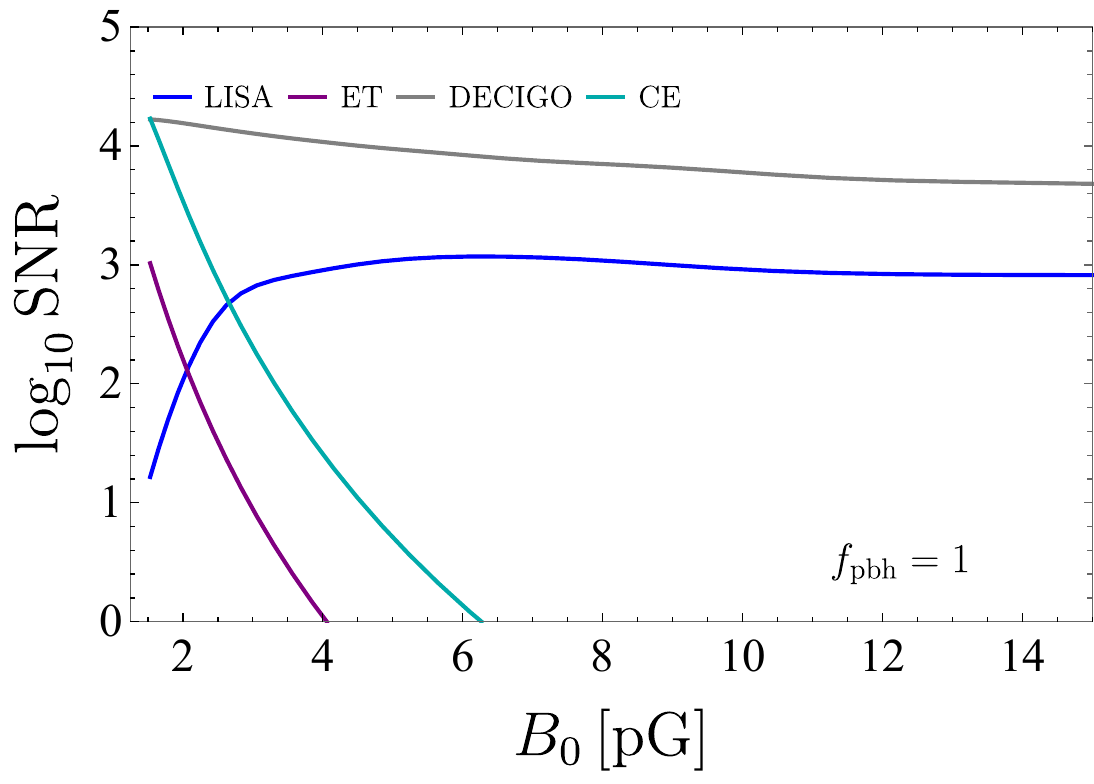}

\caption{Gravitational wave signal-to-noise ratio as a function of the magnetic field peak value in the present universe. We show the signal-to-noise ratio for LISA (blue), ET (purple), DECIGO (gray) and CE (cyan).} We consider a reheating temperature of $T_\textrm{reh} \in [10^4,10^7]$ GeV, $b=0$ and $f_\textrm{pbh}=1$. 
\label{fig:SNRB0}
\end{figure}

In Fig. \ref{fig:SNRB0}, we show the signal-to-noise ratio from GWs produced by a supercooled phase transition against the peak magnetic field amplitude. We use the respective sensitivity curves for the LISA, ET, DECIGO and CE high frequency GW experiments described in Sec.~\ref{sec:GWs} along with those predicted by the supercooled transition using \eqref{eq:snr} to calculate the signal-to-noise ratio. Future runs of experiments such as LIGO have a lower sensitivity than other experiments that cover the same subset of the frequency range such as ET, hence the prospects for the latter are more relevant. Remarkably, we see that the signal-to-noise ratios are around $\simeq 4000$ for DECIGO and span between around 1000-3000 for LISA. For ET, the signal-to-noise ratio is around 3000 for a magnetic field of 2 pG, but drops to unity at 4 pG. For CE, the signal-to-noise ratio is around 4000 for a magnetic field of 2 pG, but drops to unity around 6 pG.  We therefore observe the potential for a tantalizing multi-messenger signal yielding detectable GWs whilst producing a correlated cosmological magnetic field and all the requisite DM content of the universe in the form of PBHs.

\subsection{A model example: the classically conformal $B-L$ model}
\label{sec:b-lmodel}
To provide a concrete realization, we consider electroweak symmetry breaking in the classically scale-invariant $U(1)_{B-L}$ extension of the SM. The radiative breaking of the $B-L$ symmetry induces a negative mass term for the EW Higgs field via a portal coupling, thus eventually breaking EW symmetry. It is during this process that magnetic fields can be generated. Subsequently, plasma effects restore the EW symmetry and, finally, the universe cools down  post-reheating and the EW symmetry breaks in a similar fashion to the SM. For details on the model and its finite temperature treatment in cosmological phase transitions, see e.g.~Ref.~\cite{Baldes:2023rqv}.
Considering a $Z'$ with a mass of $m_{Z'}=56.64 \TeV$ and $B-L$ gauge coupling of $g_{B-L}=0.2832$: we get a nucleation temperature of $T_{\rm nuc}=2.03 \GeV$, a reheating temperature of $T_{\rm reh}=5.03 \TeV$, and the inverse duration of the transition $\beta/H=6.81$. In this context, the predicted PBH abundance and PBH mass are
\begin{align}
&f_{\rm PBH} = 1, \\
&M_{\rm PBH} = 5 \times 10^{-10} M_{\odot}.
\end{align}
At the same time, the peak strength of the generated magnetic fields is $B_{0}=1.9\times10^{-11} \, \textrm{G}$ at a peak coherence length of $\lambda_{0}=0.008 \, \textrm{Mpc}$. In this case we have estimated $\kappa_h$ as $\kappa_h = \Delta V_h/\Delta V_\varphi$, where $\Delta V_\varphi = V(v_\varphi,0)-V(0,0)$ and $V_\varphi = V(v_\varphi,0)-V(v_\varphi,v_h)$. $\kappa_h$ can be understood as the ratio between the heights of the scalar potential in each of the two field directions. Including the effect of $\kappa_h\simeq 4.8\times 10^{-9}$ yields a magnetic field strength of $B_0=1.3 \times 10^{-15} \, \rm{G}$ at a coherence length scale $\lambda_{0}=0.008 \, \textrm{Mpc}$.

\vspace{1.2cm}

\section{Conclusion}
\label{sec:conclusion}
We investigate the intriguing connection between strongly supercooled first-order phase transitions (FOPTs), primordial black hole (PBH) dark matter (DM) generation, and the simultaneous production of primordial magnetic fields with observable consequences. Assuming a FOPT that saturates the DM content of the universe through PBHs, we explore the resulting properties of primordial magnetic fields, including their strength and coherence length. 

We compute the universe's peak magnetic field amplitude and its coherence length, considering reheating temperatures within the range of $10^4$–$10^7$ GeV and PBH DM masses in the range $[10^{-16},10^{-10}]M_\odot$. We observe a narrow window of coherence length scales, approximately 0.001 to 0.01 Mpc, producing magnetic field amplitudes ranging from around 1 to 15 pG (assuming maximally helical magnetic fields). This easily exceeds lower limits on intergalactic magnetic fields from blazar observations.

Finally, we compute the signal-to-noise ratio from GWs produced by a supercooled phase transition against the peak magnetic field amplitude. Calculations using sensitivity curves for various GW experiments demonstrate substantial signal-to-noise ratios, reaching around $\gtrsim 10^3$ for Deci-hertz Interferometer Gravitational wave Observatory, Laser Interferometer Space Antenna, Einstein Telescope and Cosmic Explorer. This presents the exciting prospect of a multi-messenger signal, featuring detectable GWs alongside the correlated generation of cosmological magnetic fields and the formation of PBHs as the dominant DM component in the universe.

\section*{Acknowledgments}
MOOR thanks I. Baldes, K. Kamada and T. Vachaspati while MF thanks V. Vaskonen for useful discussions.    SB and MF are supported by the STFC under grant ST/X000753/1. MOOR is supported by the European Union's
Horizon 2020 research and innovation programme under grant agreement No 101002846, ERC CoG ``CosmoChart''. 

\bibliography{references}
\bibliographystyle{apsrev4-1}

\end{document}